%% file: mainprlv3.tex
\documentclass[reprint,preprintnumbers, superscriptaddress, nofootinbib
,aps,prl]{revtex4-2}

\pdfoutput=1
\usepackage{mathtools, amsfonts, amsthm, latexsym, amssymb,dsfont}
\usepackage[T1]{fontenc}
\usepackage{microtype}
\usepackage{times}
\usepackage{slashed}
\usepackage{hyperref}
\hypersetup{colorlinks=true, citecolor=blue, urlcolor=blue}
\usepackage{lipsum}
\usepackage{amsmath}
\usepackage{tikz}
\usepackage{subcaption}
\usepackage{bbm}
\newcommand{\one}{\mathbbm{1}}
\usepackage{color}
\usepackage{soul}
\newcommand{\cM}{{\mathcal M}}
\usepackage{physics}

\newcommand{\cL}{{\mathcal L}}
\newcommand{\cP}{{\mathcal P}}
\newcommand{\cV}{{\mathcal V}}

\newcommand{\cN}{{\mathcal N}}

\newcommand{\cT}{{\mathcal T}}

\newcommand{\cH}{{\mathcal H}}

\newcommand{\bZ}{{\mathbb Z}}
\newcommand{\ri}{{\rm i}}

\def\be{\begin{equation}}
\def\ee{\end{equation}}
\def\comma{\,,}
\def\period{\,.}

\def\({\left(}
\def\){\right)}

\usepackage{braket}
\usepackage{subcaption}

\makeatletter 
    
\renewcommand\onecolumngrid{
\do@columngrid{one}{\@ne}%
\def\set@footnotewidth{\onecolumngrid}
\def\footnoterule{\kern-6pt\hrule width 1.5in\kern6pt}%
}

\renewcommand\twocolumngrid{
        \def\footnoterule{
        \dimen@\skip\footins\divide\dimen@\thr@@
        \kern-\dimen@\hrule width.5in\kern\dimen@}
        \do@columngrid{mlt}{\tw@}
}%

\makeatother

\begin{document}
\title{Minimal Models RG flows:   non-invertible symmetries \& non-perturbative description}

\author{Federico Ambrosino}
\email{federicoambrosino25@gmail.com}      
\affiliation{Deutsches Elektronen-Synchrotron DESY, Notkestr. 85, 22607 Hamburg, Germany }                                                              
\author{Stefano Negro}
\email{stefano.negro@york.ac.uk}
\affiliation{Department of Mathematics, University of York,
Heslington, York YO10 5DD, UK}

\date{\today}
                
\begin{abstract}
In this letter we continue the investigation of RG flows between  {Virasoro} minimal models  {of two dimensional conformal field theories} that are protected by non-invertible symmetries. RG flows leaving unbroken a subcategory of non-invertible symmetries are associated with anomaly-matching conditions that we employ systematically to map the space of flows between minimal models beyond the $\bZ_2$-symmetric proposed recently in the literature. 
We introduce a family of non-linear integral equations that appear to encode the exact finite-size, ground-state energies of these flows, including non-integrable cases, such as the recently proposed $\mathcal{M}(k q + I,q) \to \mathcal{M}(k q - I,q)$. Our family of NLIEs encompasses and generalises the integrable flows known in the literature: $\phi_{(1,3)}$, $\phi_{(1,5)}$, $\phi_{(1,2)}$ and $\phi_{(2,1)}$.  This work uncovers a new interplay between exact solvability and non-invertible symmetries. Furthermore, our non-perturbative description provides a non-trivial test for all the  flows conjectured by anomaly matching conditions, but  so far not-observed by other means. 
\end{abstract}

\preprint{DESY-25-003}

\maketitle

\paragraph*{\textbf{Introduction.}} 
The systematic identification of Renormalization Group (RG) flows between quantum field theories is a paramount problem in theoretical physics. Global symmetries are central to this quest, providing non-perturbative constraints on the RG flows between the ultra-violet (UV) and infra-red (IR) fixed points, and dictating the allowed interactions generated along the flows. By matching their anomalies, we can put strong constraints on the IR theory. Recently, building on the seminal paper \cite{Gaiotto:2014kfa}, a profound effort has been devoted to exploring generalizations to the usual notion of global symmetries, such as \textit{higher-form}, \textit{non-invertible}, or more general \textit{higher-categorical} symmetries, extending the usual \textit{group-like} structures
to the 
more general algebraic ones of fusion higher-categories (for recent reviews see \cite{Schafer-Nameki:2023jdn, Shao:2023gho}). 
In two-dimensional conformal field theories (CFT), non-invertible symmetries are ubiquitous \cite{Chang:2018iay,Frohlich:2004ef, Fuchs:2007tx,Aasen:2020jwb}: 
topological line operators, acting as generators of $0$-form symmetries, do not form generically a group but a fusion category.  In the 
case of Rational 2d CFTs with diagonal modular invariance, i.e.\ the Virasoro Minimal Models $\mathcal{M}(p,q)$, the set of topological line operators 
coincides with the finitely many Verlinde line defect, forming a \textit{fusion modular category} \cite{Chang:2018iay,Moore:1988qv}.
Hence, the study of RG flows from a minimal model provides a unique arena where we have a complete understanding of the full set of \textit{categorical} symmetries of the UV theory, and it has indeed recently received considerable attention \cite{Benedetti:2024utz, Delouche:2024tjf,Katsevich:2024sov, Katsevich:2024jgq,Antunes:2024mfb, Antunes:2022vtb, Kikuchi:2022ipr, Kikuchi:2024cjd}. This approach was first undertaken in \cite{Chang:2018iay, Aasen:2020jwb} and, more recently, in \cite{Tanaka:2024igj}, where the authors predict infinitely many new RG flows between minimal models: $\mathcal{M}(k q + I,q) \to \mathcal{M}(k q - I,q)$, preserving a special $A_{q-1}$ fusion category containing the standard $\bZ_2$ symmetry.\\
Some specific deformations of minimal models are \emph{integrable}, meaning the scattering events
are factorized and the 2-body S-Matrix satisfies the Yang-Baxter equation 
\cite{Zamolodchikov:1978xm,Fateev:1990hy,Dorey:1996gd}. Integrable flows allow for an exact, non-perturbative description through the Thermodynamic Bethe Ansatz (TBA) equations \cite{Zamolodchikov:1989cf} -- equivalently, a Non-Linear Integral Equation (NLIE) \cite{Destri:1994bv,Klumper:1991jda} -- encoding their exact, finite-size energy spectrum.\\
Here, we extend the investigation of RG flows between minimal models predicted by anomaly-matching conditions associated with non-invertible symmetries. We also present evidence that the ground-state energy of all these RG flows -- not just the integrable ones -- admits an explicit NLIE description. We base this statement on the observation that a three-parameter family of NLIE encodes non-trivial features of the RG flows predicted by anomaly-matching conditions. In particular, for multi-operator deformations -- which is the case for most of the RG flows we looked at -- the scaling function obtained from the NLIEs shows clear signs of multi-scale behaviour in the UV, with exponents agreeing with the operators predicted by the anomaly-matching conditions. Additionally, for the deformations triggered by the $\phi_{(1,3)}$, $\phi_{(1,5)}$, $\phi_{(2,1)}$ and $\phi_{(1,2)}$ operators, the kernels of the NLIEs reduce to the known ones \cite{Martins:1992ht,Martins:1992yk,Zamolodchikov:1994za,Fioravanti:1996rz,Feverati:1999sr,Fendley:1993wq,Fendley:1993xa,Dorey:2000zb}.  {This} is substantial evidence supporting the interpretation of the NLIEs as a universal description for RG flows between minimal models. Definitive evidence will come from accurate numerical investigation and by comparison against Conformal Perturbation Theory \cite{Zamolodchikov:1989cf} or Hamiltonian Truncation \cite{Yurov:1989yu}. We will embark on this project in the near future. The NLIEs are a potential non-perturbative description for all the flows $\mathcal{M}(k q + I,q) \xrightarrow{ \phi_{(1,2k+1)}} \mathcal{M}(k q - I,q)$ conjectured in \cite{Tanaka:2024igj}, providing an explicit proof of their existence. They greatly expand the class of flows that can be studied non-perturbatively, even in the absence of a known integrable structure.\\


\paragraph{\textbf{Minimal models RG flows.}}
Consider a UV fixed point described by a Virasoro minimal model $\cT_{\rm UV} = \cM{(p,q)}$. Basic notions about Virasoro Minimal Models, fixing the conventions used in this letter, may be found in the Supplemental Material in Appendix A.
In this letter, we are interested in studying deformations of the UV theory by one of its relevant ($h_{(r,s)} <1 $) primary fields:
\be 
\mathcal{H}_{\cM{(p,q)}} + g_{(r,s)} \int\dd{x} \phi_{(r,s)}\comma
\ee
 {where $\mathcal{H}_{\cM{(p,q)}}$ is the Hamiltonian of the minimal model $\cM{(p,q)}$.}
The IR fixed point at the end of this RG flow may be either \textit{gapped} or \textit{gapless}. The former case is typical for generic deformations, having an IR described by a Topological Quantum Field Theory (TQFT).  We will not consider these in our analysis, albeit they can be studied with similar techniques to those discussed here \cite{Copetti:2024dcz,Copetti:2024rqj, Tanaka:2024igj, Chang:2018iay, Cordova:2024nux,Cordova:2024iti, Aasen:2020jwb}. We refer the reader to the discussion at the end of this letter for comments on this matter. 
In the latter case, when the IR theory is gapless,  we assume here it may be described by another minimal model itself : \begin{equation}
\label{candth}
    \mathcal{T}_{\rm UV }\,\quad = \quad   \cM{(p,q)}\xrightarrow{\phi_{(r,s)}} \cM{(p',q')} \quad = \quad \cT_{\rm IR}\period
\end{equation}  
An important constraint on $\cT_{\rm IR}$ is given by the $c_{\rm eff}$-theorem: along RG flows between $\cP\cT$-symmetric non-unitary CFT, the effective central charge:
\be 
c_{\rm eff}(p,q) = 1- \frac{6}{pq} \label{eq:ceff_th}
\ee
is monotonically decreasing \cite{Castro-Alvaredo:2017udm}, and reduces to the usual Zamolodchikov $c$-theorem \cite{Zamolodchikov:1986gt} for the case of unitary CFT. We will assume that $\cP\cT$-symmetry is always preserved along our flows, as  tested by now in all the examples considered in the literature \cite{Lencses:2022ira,Lencses:2023evr,Delouche:2024tjf,Katsevich:2024jgq,Katsevich:2024sov}, where it has been observed that CFT transition happens precisely at the spontaneous $\cP\cT$ breaking locus. Furthermore, $\cP\cT$-symmetry guarantees the reality of the energy spectrum at finite volume (and therefore of conformal dimensions) along the entire flow down to the IR CFT, as is the case for the non-unitary minimal models.  
Stringent constraints follow from the non-invertible symmetry lines of the minimal models. We will describe a very general strategy we plan to employ also for more general UV fixed points in future work. \\
Whenever, for any state on the cylinder $\ket{\Phi}$, the line $\cL_\sigma$ commutes with the deformation triggering the RG flow:
\be \label{commutator}
\left[\cL_\sigma , \phi_{(r,s)} \right] \ket{\Phi} = 0\comma
\ee 
 {then} the line operator $\cL_\sigma$ is unbroken by the deformation. The maximal subcategory $\{\cL_\sigma\}^{(r,s)}_{\rm UV} \subset \cV_{(p,q)}$ of Verlinde lines commuting with the deformation  {is} closed under fusion and generates the symmetry that is preserved along the RG flow. Using the fusion rules and Verlinde line action on the primary fields:
\begin{align}
    \cL_{\sigma} \ket{\phi_{\rho}} &= 
    \begin{tikzpicture}[baseline={(0,-0.5ex)}]
        \draw[ thick,red] (0,0) circle [radius=0.6cm];
        \draw[->,red,thick] (0.6,0) arc [start angle=0, end angle=120, radius=0.6cm];
        \fill (0,0) circle [radius=1pt];
            \node at (0.2, -0.2) {$\phi_{\rho}$};
         \node at (-0.9, -0) {$ \color{red} \cL_{\sigma}$};
    \end{tikzpicture} = \frac{S_{\sigma\rho}}{S_{0\rho}} \ket{\phi_\rho} \comma
\end{align} \eqref{commutator} is turned into a trigonometric equation for the label $\sigma$, at fixed $(r,s)$. 
In particular, it implies that the quantum dimension of the preserved lines is an RG flow invariant:
 \begin{align}\label{comm2}
    \begin{tikzpicture}[baseline={(0,-0.5ex)}]
        \draw[ thick,red] (0,0) circle [radius=0.6cm];
        \draw[->,red,thick] (0.6,0) arc [start angle=0, end angle=120, radius=0.6cm];
        \fill (0,0) circle [radius=1pt];
            \node at (0.2, -0.2) {$\phi_{\rho}$};
         \node at (-0.3, 0.2) {$ \color{red} \cL_{\sigma}$};
         \draw[ ultra thick,black] (0,0) circle [radius=1cm];
         \node at (1.4, 0.0) {$ \ket{\Phi}$};
    \end{tikzpicture} =
\begin{tikzpicture}[baseline={(0,-0.5ex)}]
        \draw[ thick,red] (-0.2,0) circle [radius=0.6cm];
        \draw[->,red,thick] (0.4,0) arc [start angle=0, end angle=120, radius=0.6cm];
        \fill (0.75,0) circle [radius=1pt];
            \node at (0.65, -0.2) {$\phi_{\rho}$};
         \node at (-0.3, 0.2) {$ \color{red} \cL_{\sigma}$};
         \draw[ ultra thick,black] (0,0) circle [radius=1.cm];
         \node at (1.4, -0.0) {$ \ket{\Phi}$};
    \end{tikzpicture} 
\end{align}
as well as the spin content of the defect Hilbert spaces associated with the preserved lines  $\cH_{\cL_{\sigma}}$\footnote{More precisely, the defect spin content in the IR is a subset of the UV one due to possible massive decouplings.}. These two pieces of  RG-invariant categorical data are to be considered as 't Hooft anomaly matching conditions in the realm of fusion categories. To explore the possible $\cT_{\rm IR}$ we proceed as follows:
\begin{enumerate}
    \item Given $\cT_{\rm UV}$, for any relevant primary field $\phi_{(r,s)}$ of $\cT_{\rm UV}$, we compute the fusion subcategory\footnote{Note that this implies the closure of this category under fusion.} of Verlinde lines $\{\cL_\sigma\}^{(r,s)}_{\rm UV}$ commuting with the perturbation via eq.\ \eqref{commutator}. 
    
    \item We generate a list of the possible minimal models  $\cT_{\rm IR}$ that satisfy  {$c_{\rm eff}\(\cT_{\rm UV}\) > c_{\rm eff}\(\cT_{\rm IR}\)$}. This list is always finite.
    
    \item We select among the $\cT_{\rm IR}$ determined above, only the minimal models containing a fusion subcategory $\{\cL_\rho\}_{\rm IR}$ of Verlinde lines coinciding  with $\{\cL_\sigma\}^{(r,s)}_{\rm UV}$. This means that all the quantum dimensions, fusion rules, and spins in the defect Hilbert spaces  in these two subcategories coincide  {with} $\{\cL_\sigma\}^{(r,s)}_{\rm UV}$. 
\end{enumerate}
This, for any given $(p,q)$ produces a list of candidate flows of the form \eqref{candth}
fulfilling all the anomaly-matching conditions by construction. This procedure shall be regarded as exclusive rather than inclusive, meaning  that anomaly-matching does not 
guarantee that the flow 
will dynamically exist. 
\begin{figure}
    \centering
    \includegraphics[width=1\linewidth]{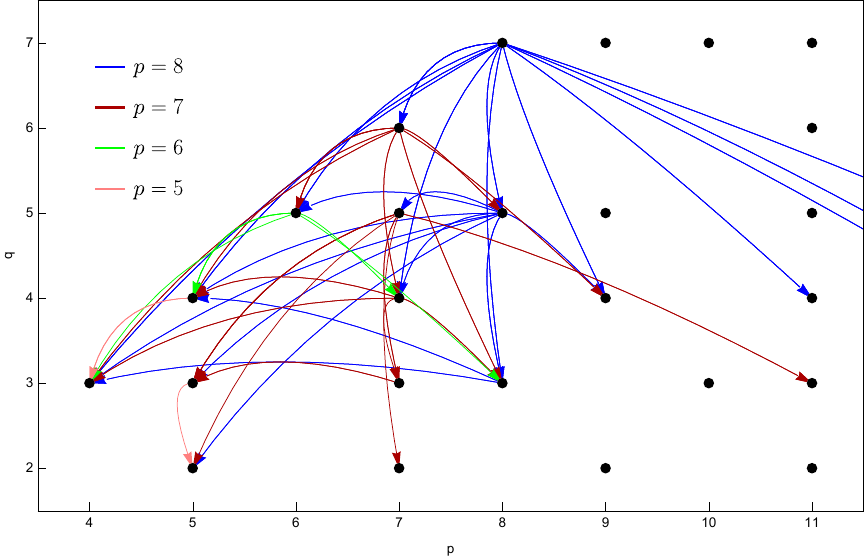}
    \caption{Flows between minimal models determined by our algorithm$\cT_{\rm UV} =\cM(p \leq 8 ,q)$.} 
    \label{minimalplot}
\end{figure}
Generically more than a single relevant operator may trigger the same flow. 
 If the set of operators triggeri {ng} the flow also preserves the same fusion subcategory  of 
lines, then along the flow all such operators ( {and all the Virasoro descendants thereof}) may be dynamically generated.
The critical point in the IR will be hit 
by fine-tuning a combination of the UV deformations, we refer to the 
appendix for a more detailed illustration. 
Lastly, the existence of a gapless flow triggered by a given relevant operator does not exclude the existence of gapped phases for other critical couplings. For example, the $\phi_{(1,3)}$ perturbation of the tricritical Ising model $\cM(5,4)$ flows to either a gapped phase or to $\cM(4,3)$ depending on the sign of the perturbation. 
 {One can also determine which fields, govern the approachentering direction of the flow to the IR.} Indeed, given that the topological lines are preserved along the flow, one can analogously determine which are the most relevant irrelevant operators of $\cT_{\rm IR}$, commuting with the same subcategory of the UV theory

 {Note that those may also be Virasoro descendants of relevant primaries.} Often the most relevant irrelevant among the operators is $T\overline{T}$,  {that is is always a viable direction given that it commutes with all the topological lines.  } In this case, the flow either enters along the $T\overline{T}$ direction\footnote{E.g.\  $\cM(3,4)\to\cM(2,5)$, or $\cM(4,5)\to \cM(3,4)$. See Appendix B for more examples.} or along the least irrelevant descendant of a (relevant) primary  that has  commutations relations with the preserved topological lines and RG invariants consistent with the UV data.  From this perspective, the non-invertible symmetries also provide a strong organizing principle 
for the effective field theory (or conformal perturbation theory) expansion around both the UV and IR. 
The procedure is easily automatized and implemented in  \texttt{Mathematica}. In Figure \ref{minimalplot} we report the outcome up to $p = 8$, but 
it can be readily extended to any value of $p$ \footnote{The known flows\cite{Lencses:2022ira,Lencses:2023evr} $\cM(2k+1,2)\to\cM(2k+1,2)$ are not displayed in the figure given that these degenerate cases where all the topological lines are broken are not determined by our method. } . Among others, we reproduce all the $\bZ_2$ symmetric flows conjectured in \cite{Tanaka:2024igj}, as well as the $\phi_{(1,2)}$, and $\phi_{(2,1)}$ flows known to be integrable \cite{Smirnov:1991uw,Dorey:2000zb}, but not belonging to that family. In addition,  {we find new flows that are not part of the known families but are allowed by anomaly matching}. An example is 
$\cM(7,5) \xrightarrow{\phi_{(2,3)}} \cM(11,3)$ discussed in the Appendix (together with a detailed discussion up to $p =7$).  An interesting case is the integrable flows with $\phi_{(1,5)}$, while $\phi_{(1,3)}$ could a priory be dynamically generated along the flow, the solutions of the NLIEs suggest that this is not the case. We plan to study in detail the relation between integrability and non-invertible symmetries somewhere else.

\paragraph*{\textbf{Description via NLIE}}
Since the seminal article \cite{Zamolodchikov:1987ti}, it was shown that certain special perturbations of minimal models could be described as \emph{quantum reductions} of integrable quantum field theories. Specifically, perturbations controlled by the relevant field $\phi_{(1,3)}$ are obtained as quantum reductions of the sine-Gordon (sG) model \cite{Smirnov:1989hh,LeClair:1989wy,Smirnov:1990vm,Martins:1992yk,Martins:1992ht}, while perturbations by the fields $\phi_{(1,5)}$, $\phi_{(2,1)}$ and $\phi_{(1,2)}$ arise from the quantum reduction of the ``Zhiber-Mikhailov-Shabat'' (ZMS) model \cite{Smirnov:1991uw,Martins:1991hv,Efthimiou:1992mx,Takacs:1996wt}. Thanks to their integrability, it has been possible to derive a Non-Linear Integral Equation (NLIE) that \emph{non-perturbatively} encodes the energies $E_{\rm s}(R)$ of any state s on a cylinder of radius $R$, which acts as an RG parameter. As functions of $r = R m$, with $m$ being the mass scale of the system, the energies interpolate between the UV regime 
$$E_{\rm s}(R)\overset{r\to 0}{\longrightarrow} - \frac{\pi(c_{\rm UV} - 24 h_{\rm s})}{6R}\comma $$ the usual Casimir behaviour \cite{Cardy:1986ie}, and the IR one 
$$E_{\rm s}(R) \overset{r\to\infty}{\longrightarrow} {\rm N}_{\rm s} \in\mathbb{Z}_{\geq 0}\period$$  In \cite{Fendley:1993wq,Fendley:1993xa,Zamolodchikov:1994za} it was shown that the integrable structure of sG could be equally well employed to encode \emph{massless flows} interpolating between successive unitary minimal models $ \cM{(p+1,p)} \xrightarrow{\phi_{(1,3)}} \cM{(p,p-1)}$. Soon it became clear that this description could also address flows 
$\cM{(p,q)}\xrightarrow{\phi_{(1,3)}}\cM{(2p-q,q)}$ \cite{Fioravanti:1996rz,Feverati:1999sr}  and, using the integrable structure of ZMS, massless flows $\cM{(2p+I,p)\xrightarrow{\phi_{(1,5)}}}\cM{(2p-I,p)}$ and lastly in \cite{Dorey:2000zb} the flows $\cM{(2p-I,p)}\xrightarrow{\phi_{(2,1)}}\cM{(2p-I,p-I)}$.

One of the main results of this letter is that the NLIEs encoding the finite size spectrum of massless flows between minimal models can be extended -- at the very least on a qualitative level -- beyond the $\phi_{(1,3)}$, $\phi_{(1,5)}$, $\phi_{(2,1)}$ and $\phi_{(1,2)}$ cases, to the whole family of flows predicted by anomaly matching conditions associated to non-invertible symmetries.

The structure of the ``massless NLIEs'' is the same as for the known cases \cite{Zamolodchikov:1994za,Dorey:2000zb}: one first computes the solutions $f_{\rm R}(\theta)$ and $f_{\rm L}(\theta)$ to the following coupled NLIE system
\begin{widetext}
\begin{equation}
    \begin{split}
        f_{\rm R}(\theta) &= \ri \alpha' - \ri \frac{r}{2} e^{\theta} - \sum_{\sigma=\pm} \sigma \intop_{\mathcal{C}_{\rm s}^{\sigma}} d\theta'\,\Big[\phi(\theta - \theta')L_{\rm R}^{-\sigma}(\theta') + \chi(\theta - \theta')L_{\rm L}^{\sigma}(\theta')\Big]\;, \\
        f_{\rm L}(\theta) &= -\ri \alpha' - \ri \frac{r}{2} e^{-\theta} + \sum_{\sigma=\pm} \sigma \intop_{\mathcal{C}_{\rm s}^{\sigma}} d\theta'\, \Big[ \phi(\theta - \theta')L_{\rm L}^{\sigma}(\theta') + \chi(\theta - \theta')L_{\rm R}^{-\sigma}(\theta') \Big]\;,
    \end{split}
    \label{eq:massless_NLIEs}
\end{equation}
%
%
%
%
\end{widetext}
where $L_{\genfrac{}{}{0pt}{3}{\rm R}{\rm L}}^{\pm}(\theta) = \log\left[1 + \exp(\pm f_{\genfrac{}{}{0pt}{3}{\rm R}{\rm L}}(\theta))\right]$. Then, the \emph{scaling function} $f_{s}(r) = 6R E_{\rm s}(R)/\pi$ is determined as
\begin{equation}
    f_{s}(r) = \sum_{\sigma=\pm} \frac{3\ri r \sigma}{2\pi^2} \intop_{\mathcal{C}_{\rm s}^{\sigma}} d\theta\, \Big[ e^{-\theta} L_{\rm L}^{\sigma}(\theta) - e^{\theta} L_{\rm R}^{-\sigma}(\theta)\Big]\;.
    \label{eq:massless_scaling_function}
\end{equation}
%
%
In these equations, the parameter $\alpha'$ is known as \emph{twist}. The kernels $\phi(\theta)$ and $\chi(\theta)$ identify the specific theory, while the contours $\mathcal{C}_{\rm s}^{\pm}$ determine the state. In particular, the ground state is obtained with the choice $\mathcal{C}_{\rm s}^{\pm} = \mathbb{R}\pm \ri \eta$, with $\eta\gtrsim 0$.

The flows described in this letter correspond to the following choice of kernels\footnote{We find it worth mentioning that these kernels were derived from an educated guess. We asked their Fourier transform to have the following properties: 1/ be a ratio of trigonometric functions; 2/ be non-vanishing at $\omega = 0$ (as this value enters the expressions of $c_{\rm eff}^{\rm UV/IR}$ and $h_{(r,s)}$); 3/ be decaying as $\omega\to\pm\infty$; 4/ present poles at $\omega=\pm\ri$ (which produce the bulk term in the UV expansion of $f(r)$); 5/ present two free parameters (that, along with the twist $\alpha$, allow to tune the UV and IR central charges and the UV conformal dimension at will). Asking additionally for a minimal number of trigonometric functions, we arrived at the expressions \eqref{eq:massless_kernels}.}
\begin{equation}
    \begin{split}
        \phi(\theta) & = - \intop_{\mathbb{R}}\frac{d\omega}{2\pi} e^{\ri \theta \omega} \frac{\sinh(\frac{1}{\kappa}\pi \omega) \cosh(\frac{2\xi - \kappa}{2\kappa} \pi \omega)}{\sinh(\frac{\xi - 1}{\kappa} \pi \omega) \cosh( \frac{1}{2} \pi \omega)}\;, \\
        \chi(\theta) & = - \intop_{\mathbb{R}}\frac{d\omega}{2\pi} e^{\ri \theta \omega} \frac{\sinh(\frac{1}{\kappa}\pi \omega) \cosh(\frac{\kappa - 2}{2\kappa} \pi \omega)}{\sinh(\frac{\xi - 1}{\kappa} \pi \omega) \cosh( \frac{1}{2} \pi \omega)}\;,
    \end{split}
    \label{eq:massless_kernels}
\end{equation}
where $\kappa > 2$ and $\xi > 1$, making the Fourier image integrable\footnote{The limit case $\kappa = 2$ yields the $\phi_{(1,3)}$ massless flows \cite{Zamolodchikov:1994za,Fioravanti:1996rz,Feverati:1999sr}. This limit is non-trivial and is discussed in details in the supplementary material.} on $\mathbb{R}$. The physical parameters of the UV and IR CFTs are determined as follows
\begin{equation}
    \begin{split}
        & c_{\rm eff}^{\rm UV}(p,q) \equiv 1 - \frac{6}{p q} = 1 - 3\(\frac{\alpha'}{\pi}\)^2 \frac{(\xi - 1)^2}{\xi(\xi +1)}\;, \\
        & c_{\rm eff}^{\rm IR}(p',q') \equiv 1 - \frac{6}{p' q'} = 1 - 3\left(\frac{\alpha'}{\pi}\right)^2 \frac{\xi - 1}{\xi}\;, \\
        & h_{(r,s)} \equiv \frac{(p r - q s)^2 - (p-q)^2}{4 p q} = 1 - \frac{1}{z_{(r,s)}}\frac{\kappa}{\xi + 1}\;.
    \end{split}
\label{eq:massless_parameter_identification}
\end{equation}
with $h_{(r,s)}$ being the conformal dimension of the perturbing field $\phi_{(r,s)}$ in the UV and $z_{(r,s)} = 1,2$ depending on whether the field $\phi_{(r,s)}$ is even or odd under the natural $\mathbb{Z}_2$ symmetry in the UV\footnote{Specifically,  {$z_{(r,s)} = 1 + \big[\big(p(r-1)-q(s-1)\big)\,({\rm mod}\,2)\big]$}.}. Fixing these three physical parameters, i.e. choosing a UV starting CFT, together with an outgoing direction, and a target IR CFT uniquely fixes the form of the NLIEs \eqref{eq:massless_NLIEs}.  Consequently, any additional information extracted from \eqref{eq:massless_scaling_function} can be considered a non-trivial prediction. One quantity that can be analytically computed is the conformal dimension of the operator that attracts the flow in the IR CFT:
\begin{equation}
    \begin{split}
        h_{(r',s')} = 1 + \frac{1}{z_{(r',s')}}\frac{\kappa}{\xi - 1}\;.
    \end{split} 
\end{equation}
The request that this conformal dimension appears, as it should, in the Ka\v{c} table of the IR minimal model $\mathcal{M}{(p',q')}$ enforces a constraint on the allowed values of the integers $p,q,p',q',r,s,r'$, and $s'$. 
\small
\begin{equation}
    \frac{p (r + 1) - q (s - 1)}{p' (r' + 1) - q' (s' - 1)} = -\frac{z_{(r',s')}}{z_{(r,s)}}\frac{p' (r' - 1) - q' (s' + 1)}{p (r - 1) - q (s + 1)}\;,
\end{equation}
\normalsize
While we could not find the most general solution to the above Diophantine equation, we can verify that the special family of solutions that corresponds to the flows discovered in \cite{Tanaka:2024igj}
\begin{equation}
    \{ \cM_{(\mu p + I,p)} \xrightarrow{\phi_{(1,2\mu + 1)}} \cM_{(\mu p - I,p)} \}\comma
    \label{eq:Nakayama_flows}
\end{equation}
solve all the constraints with $2\mu$ and $\mu p - I$ being positive integers. This family includes the familiar $\phi_{(1,3)}$, $\phi_{(1,5)}$, $\phi_{(2,1)}$ and $\phi_{(1,2)}$ flows\footnote{The case $\phi_{(2,1)}$ 
in eq. (3.9) in \cite{Dorey:2000zb} is recovered from \eqref{eq:Nakayama_flows} by setting $\mu = 1/2$, re-defining $p = 2P-J$ and $I = J/2$ and 
swapping the indices of the minimal models and of the primary fields.}. In the 
Appendix, we show how the NLIEs \eqref{eq:massless_NLIEs} reduce the known integrable cases for $\mu = 1/2,1,2$\footnote{It is perhaps worth to remark that we did not ask for the known integrable cases to be included in the educated guess that yielded \eqref{eq:massless_kernels}. This is a post-hoc result that reveals our kernels to be a $1$-parameter deformation of the known integrable ones.}.
%
%
where $p'/2\leq p\leq p'-2$. In the ancillary Mathematica notebook is included a routine that determines all the flows allowed by the above constraint. Further restrictions can be imposed on the solutions using the non-invertible symmetry matching.  

\paragraph{\textbf{Numerical analysis and conformal perturbation theory.}}
Extracting analytically any further non-trivial prediction from NLIEs of the form (\ref{eq:massless_NLIEs}, \ref{eq:massless_scaling_function}) is a notoriously arduous task. We can make some headway by studying them numerically. In particular,  we can compare the behavior of the scaling function for large and small values of $r$ to the predicted behaviour of the ground-state energy along the flow \eqref{candth}. Contrary to the well-known integrable cases, we expect the general flow to be a \emph{multi-field deformation} of the UV CFT, with the IR theory only arising upon fine-tuning of the critical coupling of the various deforming fields, e.g.\ in the flow $\mathcal{M}{(7,2)}\to\mathcal{M}{(5,2)}$, where both UV fields $\phi_{(1,2)}$ and $\phi_{(1,3)}$ were seen to contribute by using a Hamiltonian truncation method \cite{Lencses:2023evr,Lencses:2022ira}. Indeed, all relevant operators allowed by the preserved generalized symmetries will contribute to the flow, in agreement with the standard Wilsonian RG lore.  
For a flow triggered by a number $M$ of relevant UV fields $\{\phi_{(r_i,s_i)}\}_{i=1}^M$, the expected small $r$ behaviour of the scaling function \eqref{eq:massless_scaling_function} is
\begin{equation}
    \begin{split}
        f(r) \overset{r\to 0}{=} & \frac{3r^2/(4\pi)}{\sin(\frac{\pi\kappa}{\xi + 1})}
        + \sum_{\{l_i\}=0}^{\infty} a_{l_1,\ldots,l_M} r^{\sum_{i=1}^M l_i y_{(r_i,s_i)}}\comma \\
        y_{(r,s)} = & 2 z_{(r,s)}(1-h_{(r,s)})\comma \\
        a_{0,0,\ldots,0} = & c_{\rm eff}(p,q) = 1 - \frac{6}{p q}\period
    \end{split}
    \label{eq:UV_scaling_function}
\end{equation}
Here the coefficients $a_{l_1,\ldots,l_M}$ are proportional to the correlation functions of the perturbing fields on the vacuum  {(see }\cite{Klassen:1990dx}  {for more details )}. While the expansion \eqref{eq:UV_scaling_function} is expected to have a finite radius of convergence \cite{Zamolodchikov:1989cf,Klassen:1990dx}, the situation in the IR is much less under control. There, the Conformal Perturbation Theory (CPT) expansion
\begin{equation}
        f(r) \overset{r\to\infty}{=}  f(p',q') + \sum_{l=1}^\infty \Big(a_l' r^{l y_{(r',s')}} + b_l' r^{-2l}\Big) + \cdots
\label{eq:IR_scaling_function}
\end{equation}
is asymptotic, and there is 
little \cite{Zamolodchikov:1991vh} control over the omitted further contributions. 
We performed a numerical analysis of the NLIEs \eqref{eq:massless_NLIEs} for several cases and found that, in all of them, the scaling function \eqref{eq:massless_scaling_function} agrees perfectly with the expected behaviours (\ref{eq:UV_scaling_function}, \ref{eq:IR_scaling_function}). While the parameters \eqref{eq:massless_parameter_identification}, are built in the kernel by construction, the agreement with the multiple sum for small $r$ shall be regarded as a highly non-trivial check that our data passes with flying colours. In principle, further support can come from comparing the first few coefficients of the expansions with the estimates coming from CPT. We will report on this in a future publication. 
\begin{figure}
\centering\includegraphics[width=1\linewidth]{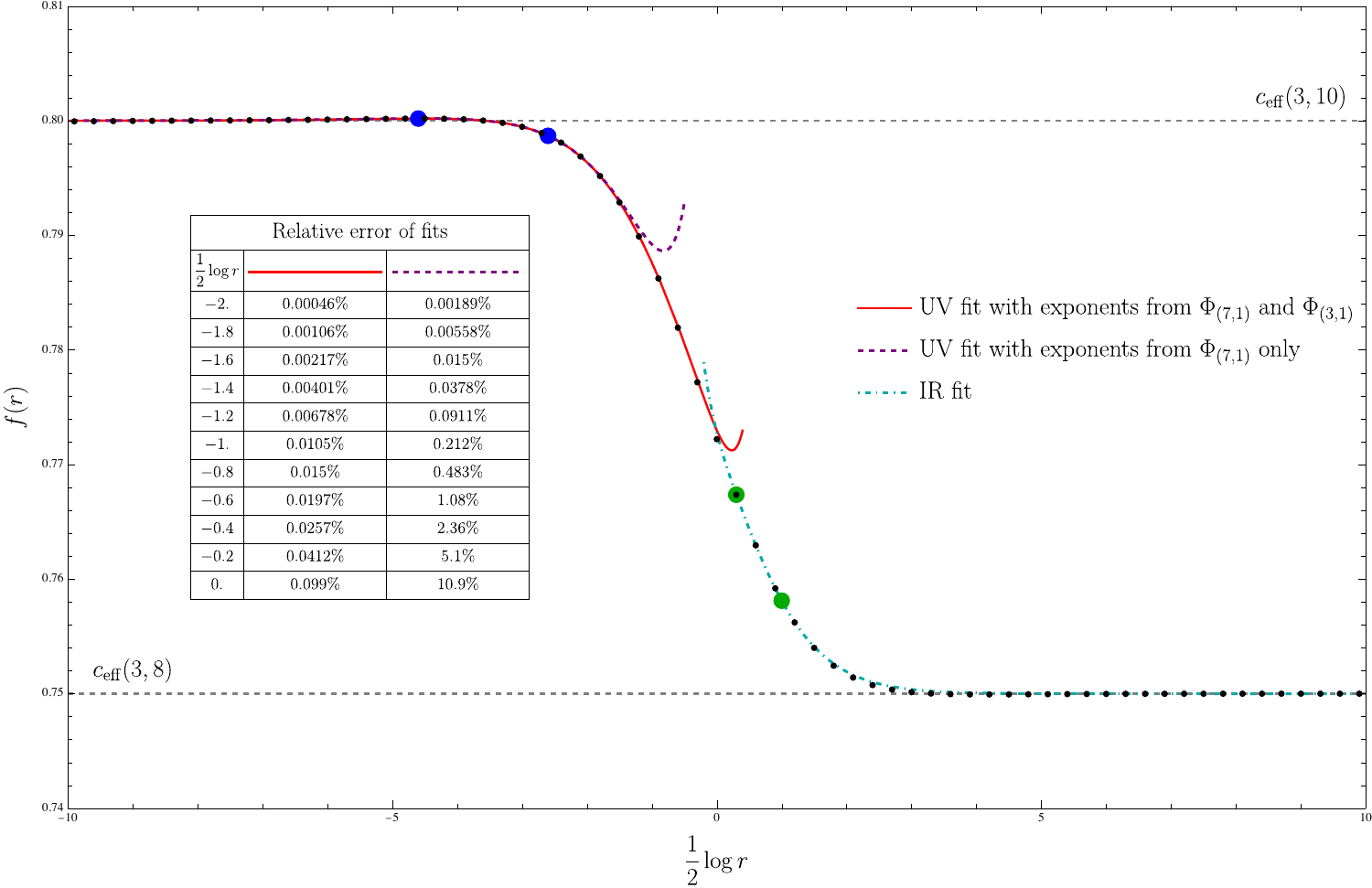}
    \caption{
    Ground state scaling function \eqref{eq:massless_scaling_function} 
    for the flow $\mathcal{M}{(10,3)} \to \mathcal{M}{(8,3)}$, triggered by $\phi_{(1,7)}$. 
    We fitted the points between the pairs of larger blue and green dots against UV and IR CPT predictions, respectively. The fit performed including the contributions of all perturbing fields (red full line) -- here $y_{(1,5)} = 3 y_{(1,7)}$ -- performs much better than the fit for a single field 
    (purple dashed line). The table collects the relative errors 
    on points that were not used 
    .}
    \label{fig:Kleb_fit}
\end{figure}
Figure \ref{fig:Kleb_fit} reports the numerical results for the flow $\mathcal{M}(10,3) \to \mathcal{M}(8,3)$, triggered by $\phi_{(1,7)}$, first proposed in \cite{Fei:2014xta}, which has recently received a lot of attention \cite{Katsevich:2024jgq,Klebanov:2022syt,Delouche:2024tjf}. The fit that includes contributions from all the perturbing fields, is numerically favoured, independently agreeing with the results obtained in \cite{Delouche:2024tjf} by employing Hamiltonian truncation and CPT methods.

\paragraph{\textbf{Outlook.}}
In this paper, we studied RG flows between  generic minimal models. Many flows can be conjectured by the matching of the global symmetries. For these flows, we propose an NLIE description encoding the ground state energy non-perturbatively. It would be interesting to confront our ground state energy with the results that can be independently obtained by  Conformal Perturbation Theory and Hamiltonian Truncation. While here we focussed on gapless RG flows between Minimal Models,  our methods extend to the ones to gapped phases. In this case, the anomalies of non-invertible symmetries predict a non-trivial structure of the vacua of the TQFT and particle-soliton degeneracies \cite{Cordova:2024nux,Cordova:2024iti,Copetti:2024dcz,Copetti:2024rqj}. A direct application  would be to check whether the RG flows  between QCD$_2$ theories proposed in \cite{Delmastro:2022prj} may be obtained via matching of the anomalies associated with lines of the coset models in QCD$_2$ as initiated recently in \cite{Cordova:2024nux}. Another interesting future direction is studying  the IR fixed point from deformation of coupled minimal models, following the approach of \cite{Antunes:2024mfb,Antunes:2022vtb, Klebanov:2022syt}, especially because these flows may end on compact non-rational CFT for some special deformations. \\ 
Our NLIEs also admit a simple extension to the massive version, similar to what happens for the $\phi_{(1,3)}$ and $\phi_{(1,5)}$, $\phi_{(1,2)}$ cases. For these integrable massive flows, the NLIE describes the ground state energy of, respectively the sG and ZMS theories. In general, we expect the massive version of our equations to be related to the ground state of a (time-like) Liouville CFT deformed by several vertex operators. This perspective suggests the possibility of studying the one-point functions of these theories using the reflection relations proposed in \cite{Lukyanov:1996jj,Fateev:1997nn,Fateev:1997yg}. We plan to follow this path in the near future. 
\paragraph{\textbf{Acknowledgments.}} We thank Patrick Dorey, Shota Komatsu, Alessio Miscioscia, Volker Schomerus, Roberto Tateo and Igor Klebanov for useful discussions. We also thank the anonymous referees for having read carefully our manuscript and for providing detailed and precious feedback.
FA gratefully acknowledges support from the Simons Center for Geometry and Physics, Princeton Center for Theoretical Science, and Perimeter Institute, at which some of the research for this paper was performed.
 FA thanks the Deutsche Forschungs
gemeinschaft (DFG, German Research Foundation) under Germany’s Excellence Strategy
EXC 2121 “Quantum Universe”- 390833306, and the Collaborative Research Center- SFB
 1624 “Higher structures, moduli spaces and integrability”- 506632645, for support.

\bibliography{Ref}

\appendix
\onecolumngrid
\newpage

\include{suppl.tex}

\end{document}

%% file: suppl.tex
\section{Appendix A: Primer on minimal models.}

In this section we collect  some basic notions on minimal models and we fix the conventions used in this letter. Virasoro Minimal Models $\cM{(p,q)}$, {with $p,q$ being coprime integers (we take  with $p>q$)}, are 2d rational CFTs enjoying diagonal Modular Invariance. Their central charge is:
\begin{equation}
    c(p,q) = 1- 6 \frac{(p-q)^2}{pq}\comma
\end{equation}
they consist of $(p-1)(q-1)/2$ primaries $\phi_{(r,s)}$ in a fundamental domain {$r=1,\cdots , q-1\comma s = 1,\cdots, p-1$ with $sq + rp < pq$}, having conformal weights:
\begin{equation}   \label{dimension}
    h_{(r,s)} = h_{(q-r,p-s)} =  \frac{(pr - qs)^2 - (p-q)^2}{4 pq}\period
\end{equation} 
For simplicity, we will often denote the primaries just by {$\phi_\rho$}, with $\phi_0 = \phi_{(1,1)} = \mathds{1}$.

The primary fields form the fusion ring:
\be 
\left[\phi_{\rho} \right] \times \left[\phi_{\sigma} \right] = \sum_{\kappa} N^{\kappa}_{\rho\sigma} \,\left[\phi_{\kappa}\right]  
\ee
where the fusion coefficients $N^{\kappa}_{\rho\sigma}$ are given in terms of the modular S-matrix {$S_{\rho\sigma}$} as:
\begin{align}
    N^\kappa_{\rho\sigma} &= \sum_\lambda \frac{S_{\rho\lambda}\,S_{\sigma\lambda} S_{\kappa\lambda}}{S_{0\lambda}}\\
        S_{(r,s),(\rho,\sigma)} &= (-1)^{1 + s\rho + r\sigma}\sqrt{\frac{8}{pq}} \sin(\pi\frac{p}{q} r \rho) \, \sin(\pi\frac{q}{p} s \sigma)
\end{align}
The topological Verlinde line operators, are in one-to-one correspondence with the primary fields $\cL_{(r,s)}$ (that we will also equivalently denote  as $\cL_\rho$). They satisfy the same fusion ring as the primary fields:
\be 
\cL_{\rho} \times \cL_{\sigma} = \sum_{\lambda} N_{\rho\sigma}^{\lambda}\, \cL_{\lambda}  \period
\ee
We denote the fusion modular category generated by the Verlinde lines by $\cV_{(p,q)}$.

Let us denote by $\phi_\rho \ket{0} = \ket{\phi_\rho}$ a state on the cylinder.  Then, the action of a Verlinde line on a primary field is given by:
\begin{align}
    \cL_{\rho} \ket{\phi_{\sigma}} &= 
    \begin{tikzpicture}[baseline={(0,-0.5ex)}]
        \draw[ thick,red] (0,0) circle [radius=0.6cm];
        \draw[->,red,thick] (0.6,0) arc [start angle=0, end angle=120, radius=0.6cm];
        \fill (0,0) circle [radius=1pt];
            \node at (0.2, -0.2) {$\phi_{\sigma}$};
         \node at (-0.9, -0) {$ \color{red} \cL_{\rho}$};
    \end{tikzpicture} = \frac{S_{\rho\sigma}}{S_{0\sigma}} \ket{\phi_\sigma} \period
\end{align}
The action of a Verlinde line on a field, should be thought as generating the Ward Identity for the discrete symmetries associated to the topological line. 
The vacuum expectation value of a Verlinde line on the cylinder is known as \textit{Quantum dimension}:\be 
\langle \cL_\rho\rangle= \bra{0}\cL_\rho \ket{0}  = 
 \begin{tikzpicture}[baseline={(0,-0.5ex)}]
        \draw[ thick,red] (0,0) circle [radius=0.6cm];
        \draw[->,red,thick] (0.6,0) arc [start angle=0, end angle=120, radius=0.6cm];
         \node at (-0.9, -0) {$ \color{red} \cL_{\rho}$};
    \end{tikzpicture} = d_\rho \period
\ee
Whenever the symmetry generated by $\cL_\rho$ is an invertible and non-anomalous group-like element, then $d_\rho =1 $, but it is { an irrational} number for a non-invertible element of the fusion category, and it can be dressed by a phase whenever the symmetry has a 't Hooft anomaly. 

The torus partition function with the insertion of a Verlinde line operator $\cL$ along the spatial cycle is after a modular transformation, the partition function  over the defect Hilbert space $\cH_{\cL}$ of local operators living at the endpoint of $\cL$. It is given explicitly by a twisted trace over the Virasoro characters:
\be \label{twistedtrace}
Z_{\cL_{\lambda}}(\tau,\overline{\tau}) = \sum_{\rho,\sigma}\, N^\lambda_{\rho\sigma} \,{\chi}_\rho({\tau}) \,\overline{\chi}_\sigma(\overline{\tau}) \period
\ee
From \eqref{twistedtrace}  we read off the spin content of the defect Hilbert space by evaluating \be s_{\lambda} = (h_\rho - h_\sigma)
 \mod\bZ\ee over non-zero fusion coefficients. Spins being non-semi integers signal the presence of a 't Hooft anomaly of the global symmetry generated by $\cL_\lambda$

\section{Appendix B: Detailed description of RG from $\boldsymbol{\cM_{(p,q)}}$ with $\boldsymbol{pq \leq 42}$}
In this appendix we provide a detailed discussion of all the flows that can be predicted based on generalized symmetries constraints from a UV fixed point $\cM(p,q)$ with  {$pq \leq 42$, corresponding to all minimal models having effective central charge less or equal than the one of $\cM(7,6)$.}\footnote{{We thank the anonymous referee of our paper for the valuable suggestion of organizing these flows by their effective central charge.}}. 

{We remind that in order to determine putative RG flows, we proceed as explained in the main text below }eq.\ (6). {Specifically, fixed the UV CFT $\cM(p,q)$:

1) for any of its relevant deformations, we compute the fusion subcategory (if there is any) that is left unbroken by the deformation; 

2) we determine all the minimal models having effective central charge less than the one of the UV theory;

3) we select only the IR models having a fusion subcategory whose RG invariants match the UV one.}

{Let us remark that the case of $\cM(2k+1,2)$ is quite degenerate as any relevant deformation breaks all the non-invertible lines}\cite{Tanaka:2024igj}{. For this reason, flows from  $\cM(2k+1,2)$ are hard to identify using this method. Of course flows between these multicritical YL models are well known in the literature}\cite{Lencses:2022ira,Lencses:2023evr}{, but we will  not discuss them here. }

We organize the flows by increasing value of $pq$ of the UV theory, or equivalently or increasing UV effective central charge. {Unless explicitly indicated, we do find a  choice of parameters $\alpha'$, $\xi$ and $\kappa$ in the NLIE kernels for all these flows. We do not report a detailed analysis for all of them here, except for the $\cM(10,3)\to \cM(8,3)$ discussed in the main text and chain of flows presented in Appendix D. }
In the following, we denote $\cL_{(1,1)}:= \mathbbm{1}$. 
\begin{itemize}
    \item \large$\boldsymbol{\cM(5,3) \xrightarrow{\phi_{(2,1)}} \cM(5,2)}$. \normalsize The relevant $\(h_{(1,2)} = \frac{3}{4}\)$ deformation preserves the non-unitary Fibonacci category\footnote{There are two {F}ibonacci categories with fusion rules $W^2 = 1+ W$, distinguished by their anomalies (F-symbols), or equivalently, by the defect Hilbert Space spins \cite{Chang:2018iay} } generated by the non-invertible element $\cL_{(1,3)}$, with \be \label{fibcat}
    d_{(1,3)} = \frac{1}{2} \left(1-\sqrt{5}\right)\comma \qquad s_{{(1,3)}} = \bZ \pm \left\{0,  \frac{1}{5}\right\}\comma\qquad \cL_{(1,3)}\,\times\, \cL_{(1,3)}  =\mathbbm{1} + \cL_{(1,3)}\period
    \ee
$\cL_{(1,3)}$ of $\cM_{3,5}$ flows to the unique non-trivial line $\cL_{(1,2)}$ of the Lee-Yang model $\cM(5,2)$ that famously satisfies the anomalous fusion category above.
    More interesting is the fate of the $\phi_{(2,1)}$ primary field that enters in the IR in the $T\overline{T}$-direction\footnote{We thank Roberto Tateo for clarifications on this point.} 
This deformation was known to be integrable from \cite{Dorey:2000zb}.
\item {\large $\boldsymbol{\cM{(5,4)} \xrightarrow{\phi_{(1,3)}} \cM{(4,3)}}$} . This is gives the famous integrable Zamolodchikov flow preserving the entire Ising fusion category  (Tambara-Yamagami TY$_2$ category)   generated by the Kramers-Wannier duality defect $\cL_{{(3,1)}} := \cN$ and the invertible $\bZ_2$-line $\cL_{(2,1)}:=\eta$. This flow is studied in  \cite{Chang:2018iay}, and we will not repeat the discussion here, apart from mentioning that the $\phi_{(1,3)}$ flow enters purely along the $T\overline{T}$ direction.  We emphasize, that, in principle, there could be a putative flow $\cM{(5,4)} \xrightarrow{\phi_{(1,2)}} \cM{(4,3)}$ only preserving the invertible  $\bZ_2$-line $\eta$ but not the Kramers-Wannier duality defect $\cN$. Such a flow could only happen if the Kramers-Wannier duality is restored by symmetry enhancement in the IR.   
   \item {\large $\boldsymbol{\cM(7,3) \xrightarrow{\phi_{(1,5)}} \cM(5,3)}$.} This flow is triggered by the most relevant field $\phi_{(1,5)}$ preserving the anomalous $\bZ_2$ line $\cL_{(2,1)}$ \cite{Cordova:2019wpi}:
   \be 
\cL_{(2,1)} \times \cL_{(2,1)} = \mathbbm{1}\comma\quad d_{(2,1)} = -1\comma\quad s_{(2,1)} = \bZ \pm \frac{1}{4}
   \ee 
   Along the flow also $\phi_{(1,3)}$ could be dynamically generated given that it commutes with $\cL_{(2,1)}$ as well. In the IR, the flow enters along the $T\bar{T}$ direction, with contribution also from the descendants of $\phi_{(1,2)}$ of the IR model. Yet, it has been observed that $\phi_{(1,3)}$ does not contribute, and $\phi_{(1,5)}$ is a single-field integrable deformation. Indeed, integrability protects further the flow, by generating this operator. We plan to go back to this point somewhere else.

   \item {\large {$\boldsymbol{\cM(8,3) \xrightarrow{\phi_{(1,5)}} \cM(4,3)}$.}} This flow is triggered by the most relevant deformation $\phi_{(1,5)}$. It is a known integrable flow preserving a non-anomalous $\bZ_2$ symmetry generated by $\{\mathds{1}, \cL_{(2,1)} \}$ in the UV and entering in the IR along the irrelevant $\phi_{(3,1)}$ direction. Also the operator $\phi_{(1,3)}$ is generated along the flow. 
      {We find that the same deformation but flowing to $\cM(5,4)$ would be also possible based on $\bZ_2$ symmetry preservation. Yet,  we only find values of the NLIE parameters that can reproduce the flow to $\cM(4,3)$ and not the one to $\cM(5,4)$.}

     \item {\large $\boldsymbol{\cM(7,4) \xrightarrow{\phi_{(2,1)}} \cM(7,3)}$.} It preserves the category generated by $\left\{\mathbbm{1} , \cL_{(1,3)}, \cL_{(1,5)}\right\}$ that is an $SU(2)_{4}$ fusion ring:
   \begin{equation}
   \begin{gathered}
        \cL_{(1,3)}\times  \cL_{(1,3)} =  \mathbbm{1} + \cL_{(1,3)} + \cL_{(1,5)}\comma \qquad \cL_{(1,5)}\times \cL_{(1,5)} = \mathbbm{1} + \cL_{(1,3)} \\
        \cL_{(1,3)} \times  \cL_{(1,5)} = \cL_{(1,5)} \times  \cL_{(1,3)} = \cL_{(1,3)} +\cL_{(1,5)}\comma 
        \end{gathered}
        \end{equation} and anomalies:
        \begin{equation}\begin{aligned}
        &d_{(1,3)} = -\frac{1}{2} \csc \left(\frac{3 \pi }{14}\right)\comma  \qquad   &&s_{(1,3)} = \bZ \pm \left\{ 0  \comma  \frac{1}{7}\comma\frac{2}{7} \right\}\comma\\
      &d_{(1,5)} = 2 \sin \left(\frac{\pi }{14}\right) \comma \qquad &&s_{(1,5)} = \bZ \pm \left\{ 0\comma  \frac{2}{7}\comma\frac{5}{7} \right\}\period
         \end{aligned}
   \end{equation} 
   This flow is known to be integrable from \cite{Dorey:2000zb}. No other operators commute with this fusion category, and no other operator (apart from its descendants) may be generated along the flow.  The flow enters in $\cM(3,7)$ along the $\phi_{(2,1)}$. The line operators $\cL_{(1,3)}, \cL_{(1,5)}$ flow to the same line operators of $\cM(7,3)$.
   \item {\large $\boldsymbol{\cM(7,4) \xrightarrow{\phi_{(1,4)}} \cM(5,4)}$.} The flow is generated by $\phi_{(1,4)}$ and preserves only the invertible non-anomalous $\bZ_2$ line $\cL_{(3,1)}$, but breaks the KW duality defect. In this flow also the relevant operator $\phi_{(1,2)}$ may be generated, that is also dynamically generated. The flow enters along the $\phi_{(1,4)}$ direction. It naturally implies all the flows from $\cM(5,4)$. 
   \item {\large  $\boldsymbol{ \cM(7,4) \xrightarrow{\phi_{(1,4)}} \cM(8,3)}$.} This flow is generated by $\phi_{(1,4)}$ and $\phi_{(1,2)}$ with the former being the most relevant.  It enters along the $\phi_{(2,1)}$ direction. {We do not find any values of the NLIE parameters that agree with this flow.}
   \item {\large {$\boldsymbol{\cM(10,3) \xrightarrow{\phi_{(1,7)}} \cM(8,3)}$.}} This flow is triggered by the most relevant deformation $\phi_{(1,7)}$. $\phi_{(1,3)}$ and  $\phi_{(1,5)}$ are also generated along the flow \cite{Delouche:2024tjf,Klebanov:2022syt} and it preserves a non-anomalous $\bZ_2$ symmetry generated by $\{\mathds{1}, \cL_{(2,1)} \}$ in the UV and entering in the IR along the irrelevant $\phi_{(2,1)}$ direction. We note that the same deformation would allow for a flow to $\cM(7,4)$ with the same choice of deformation. 
We also mention that there are clearly also chained flows from $\cM(10,3)\to \cM(8,3)\to
\cM(5,4)\to \cM(4,3)$ consistent with the flows already presented.
   
    \item   { \large$\boldsymbol{\cM(6,5) \xrightarrow{\phi_{(1,3)}} \cM(5,4)}$}. This is the integrable flow between tetracritical Ising and tricritical Ising, this is discussed in details in  \cite{Chang:2018iay}. The flow preserves the non-invertible lines of $\{\cL_{{2,1}},\cL_{{3,1}},\cL_{{4,1}} \}$ flowing to the  $\{\cL_{{1,2}},\cL_{{1,3}},\cL_{{3,1}} \}$ lines in $\cM(5,4)$ respectively, having spin content:
    \be 
    s_{(2,1)} = \bZ\pm \left\{0\comma \frac{1}{10}\comma \frac{2}{5}\comma \frac{1}{2} \right\} \comma \quad  s_{(3,1)} = \bZ\pm \left\{0\comma\frac{2}{5} \right\} \comma \quad  s_{(4,1)} = \bZ\pm \left\{0\comma\frac{1}{2} \right\} \period
    \ee
    The flow enters along the $\phi_{3,1}$ direction in the IR. We also see the corresponding flow to $\cM(4,3)$.
    
    \item {\large $\boldsymbol{\cM(6,5) \xrightarrow{\phi_{(2,3)}} \cM(7,4)}$.} The flow is allowed by the effective central charge theorem as $c_{\rm eff}\( 7,4\)= \frac{11}{14}$, $c_{\rm eff}\( 6,5\)= \frac{5}{6}$. The $\phi_{(2,3)}$ deformation preserves the non-anomalous $\bZ_2$ generated by the $\cL_{(4,1)}$ line, but breaks the KW duality defects. In the IR, the flow enters along the $T\overline{T}$ direction.
    \item {\large $\boldsymbol{\cM(6,5) \xrightarrow{\phi_{(2,3)}} \cM(8,3)}$.} With the same field, there could also be a flow to this minimal model preserving the same symmetry line.  It could enter along the first descendent of $\phi_{(1,5)}$. This is also consistent with the flow $\cM(7,4) \to \cM(8,3)$ also predicted above. 
    {We do not find any values of the NLIE parameters that can agree with this flow.}

\item {\large {$\boldsymbol{\cM(11,3) \xrightarrow{\phi_{(1,7)}} \cM(7,3)}$.}}
This is a flow generating also $\phi_{(1,5)}$ and $\phi_{(1,3)}$ and preserving an anomalous $\bZ_2$ symmetry with $d_{(2,1)} = -1, s_{(2,1)} = \bZ \pm \frac{1}{4}$. It enters along the direction of descendants of $\phi_{(2,2)}$.
Clearly the same deformations also allow a flow to $\cM{(5,3)}$  by tuning of the critical couplings as consistent with the $\cM{(7,3)}\to \cM{(5,3)}$ flow discussed above.

\item {\large $\boldsymbol{\cM(7,5) \xrightarrow{\phi_{(2,1)}} \cM(7,2)}$.} This is triggered by the relevant field $\phi_{(2,1)}$ and preserves the $SU(2)_4$ category generated by the same lines as in $\cM(7,4) \to \cM(7,3)$ but with:
\begin{equation}
    \begin{aligned}
    &d_{(1,3)} = \sin \left(\frac{\pi }{7}\right) \sec \left(\frac{3 \pi }{14}\right) \comma  \qquad   &&s_{(1,3)} = \bZ \pm \left\{ 0  \comma  \frac{1}{7}\comma\frac{3}{7} \right\}\comma \\
      &d_{(1,5)} = -2 \sin \left(\frac{3 \pi }{14}\right) \comma \qquad &&s_{(1,5)} = \bZ \pm \left\{ 0\comma  \frac{1}{7}\comma\frac{2}{7} \right\}\period
    \end{aligned}
\end{equation}
It enters along the $T\bar{T}$ direction.
\item {\large $\boldsymbol{\cM(7,5) \xrightarrow{\phi_{(1,3)}} \cM(5,3)}$.} The flow triggered by $\phi_{(1,3)}$ preserves the category $\{\one\comma \cL_{(2,1)}\comma\cL_{(3,1)}\comma \cL_{(4,1)} \}$, satisfying the fusion ring:
\begin{equation}
\begin{aligned}
   &\cL_{(2,1)}\times \cL_{(2,1)} = \one + \cL_{(3,1)}\comma \quad &&\cL_{(2,1)} \times \cL_{(3,1)} = \cL_{(2,1)} + \cL_{(4,1)}\comma \quad    &&\cL_{(2,1)} \times \cL_{(4,1)} =   \cL_{(3,1)}\\
     &\cL_{(3,1)}\times \cL_{(2,1)} = \cL_{(2,1)} + \cL_{(4,1)}\comma \quad &&\cL_{(3,1)} \times \cL_{(3,1)} = \one + \cL_{(3,1)}\comma \quad    &&\cL_{(3,1)} \times \cL_{(4,1)} =   \cL_{(2,1)}\\
         &\cL_{(4,1)}\times \cL_{(2,1)} = \cL_{(3,1)} \comma \quad &&\cL_{(4,1)} \times \cL_{(3,1)} =  \cL_{(2,1)}\comma \quad    &&\cL_{(4,1)} \times \cL_{(4,1)} =   \one\comma
   \end{aligned}
\end{equation}
and RG invariants:
\begin{equation}
\begin{aligned}
&d_{(2,1)} = \frac{-1}{2} (1 + \sqrt{5})\comma \qquad     &&d_{(3,1)} = \frac{1}{2} (1 + \sqrt{5})\comma\qquad && d_{(4,1)} = -1\comma\\
&s_{(2,1)} = \bZ \pm \left\{ \frac{1}{20}\comma  \frac{1}{4}\comma \frac{9}{20}\right\}\comma \quad && s_{(3,1)} = \bZ \pm \left\{ 0 \comma  \frac{1}{5} \right\}\comma \quad && s_{(4,1)} = \bZ\pm \frac{1}{4} \period
\end{aligned}\end{equation}
The flow enters along the $T\bar{T}$ direction in $\cM(5,3)$.  Further primaries cannot be generated along the $\phi_{(1,3)}$ flow. $\phi_{(1,2)}$ would only preserve $\cL_{(3,1)}$, and the relevant fields $\phi_{(2,2)},\phi_{(2,4)}$ would only preserve $\cL_{(4,1)}$ and are therefore not allowed. We also see the corresponding tuned $\cM(7,5)\to \cM(5,3)\to \cM(5,2)$.
\item  {\large $\boldsymbol{\cM(7,5) \xrightarrow{\phi_{(2,2)}} \cM(11,3)}$.} This flow along may generate also the less relevant $\phi_{(2,4)}$  preserves the anomalous $\bZ_2$ symmetry generated by $\cL_{(4,1)}$, with \begin{equation}
    d_{(4,1)} = -1\comma \qquad s_{(4,1)} = \bZ \pm \frac{1}{4}\;.
\end{equation}  It enters in the IR along the direction of $\phi_{(2,2)}$ which preserves $\cL_{(2,1)}$ (being the image of $\cL_{(4,1)}$ of the UV).  We also see a flow to $\cM(7,3)$ through $\cM(7,5)\to \cM(11,3)\to \cM (7,3)$

\item {\large {$\boldsymbol{\cM(9,4) \xrightarrow{\phi_{(1,5)}} \cM(7,4)}$.}}
This flow, triggered by $\phi_{(1,5)}$ also generates $\phi_{(1,3)}$. It preserves the TY$_2$ category generated by ${\mathds{1},\cL_{(2,1)},\cL_{(3,1)}}$ $d_{(2,1)} = -\sqrt{2}, d_{(3,1)}=1$ and $s_{(2,1)} = \bZ \pm \left\{ \frac{3}{16}, \frac{5}{16}\right\}$ and entering along the $\phi_{(2,1)}$ direction in the IR. We also find all the flows to the various end points of the following chain: $\cM(9,4)\to \cM(7,4) \to \cM(8,3)\to \cM(5,4)\to \cM(4,3)$. 
 \item {\large {$\boldsymbol{\cM(9,4) \xrightarrow{\phi_{(1,4)}} \cM(6,5)}$.}}
 This flow, allowing also for $\phi_{(1,2)}$ to be generated, only commutes with the not-anomalous $\bZ_2$ symmetry, and enters along the $\phi_{(1,4)}$ direction in the IR. {In this case, we do find values of the NLIE parameters that agree with the flow, however, they identify $T\overline{T}$ as the operator controlling the incoming direction of the flow in the IR.}

\item {\large {$\boldsymbol{\cM(13,3) \xrightarrow{\phi_{(1,9)}} \cM(11,3)}$.}}
This flow, triggered by $\phi_{(1,9)}$ allows for all the other $\phi_{(1,2k+1)}$ to be generated and commute with the anomalous $\bZ_2$ generated by ${\mathds{1},\cL_{(2,1)}}$ and as already described above for the flows from $\cM(11,2)$, and enters along the $\phi_{(9,1)}$ direction in the IR. We also find all the flows to the various end points of the following chain: $\cM(13,3)\to \cM(11,3) \to \cM(7,3)\to \cM(5,3)$. Anomaly matching also predict a flow to $\cM(7,5)$ with the same deformation.

\item {\large {$\boldsymbol{\cM(8,5) \xrightarrow{\phi_{(2,3)}} \cM(9,4)}$.}} This flows preserves the non-anomalous $\bZ_2$ symmetry and enters along the second descendant of $\phi_{(1,4)}$ direction in the IR. We also see consistently  the flows $\cM(8,5)\to \cM(9,4)\to\cM(7,4)\to \cM(5,4)$ 

\item {\large {$\boldsymbol{\cM(8,5) \xrightarrow{\phi_{(1,4)}} \cM(7,5)}$.}} This flow preserves the non-anomalous $\bZ_2$ symmetry as above. It enters in the IR along the $\phi_{(1,4)}$ direction.

\item {\large $\boldsymbol{\cM(7,6) \xrightarrow{\phi_{(1,3)}} \cM(6,5)}$.} This is one the flow of the Ising series. It is integrable with deformation $\phi_{(1,3)}$ that preserves the whole subcategory $\left\{\cL_{(1,1)}\comma \cL_{(2,1)}\comma\cdots\comma  \cL_{(5,1)}\right\}$. For the sake of clarity, let us denote $\cL_{(k,1)}$ as $(k)$. Then the fusion ring is:
\begin{equation}
    \begin{aligned}
        &(2) \times (2) = \one + (3)\comma  &&(2)\times (3) = (2) + (4)\comma  &&(2)\times (4) = (3) + (5)\comma   &&(2)\times (5) = (4)\\
        &(3) \times (2) = (2) + (4)\comma  &&(3)\times (3) = \one + (3) + (5)\comma  &&(3)\times (4) = (2) + (4)\comma   &&(3)\times (5) = (3)\\
        &(4) \times (2) = (3) + (5)\comma  &&(4)\times (3) =  (2) + (4)\comma  &&(4)\times (4) = \one + (3)\comma   &&(4)\times (5) = (2)\\
        &(5) \times (2) = (4)\comma  &&(5)\times (3) = (3)\comma  &&(5)\times (4) = (2)\comma   &&(5)\times (5) = \one  \comma
    \end{aligned}
\end{equation}
with anomalies:
\begin{equation}
\begin{aligned}
    &d_{(2,1)} = \sqrt{3}\comma  &&d_{(3,1)} = 2\comma  &&d_{(4,1)} = \sqrt{3}\comma &&d_{(5,1)} = 1\\
    &s_{(2,1)} = \bZ \pm \left\{\frac{1}{24}\comma\frac{1}{8}\comma\frac{3}{8}\comma\frac{11}{24}     \right\}\comma  &&s_{(3,1)} = \bZ \pm \left\{0,\frac{1}{3}\comma\frac{1}{2}\right\}\comma  &&s_{(4,1)} = s_{(2,1)}\comma &&s_{(5,1)} = \bZ \pm\left\{0,\frac{1}{2}\right\}\\
\end{aligned}
\end{equation}
It is integrable with $\phi_{(1,3)}$. Deformations by $\phi_{(2,1)},\phi_{(3,2)},\phi_{(3,3)}$ would break part of this global symmetry. As argued above, it is very unlikely that these flow would have $\cM(6,5)$ as an IR fixed point given that the broken symmetries would need to be restored at the end of the flow. {Hence,} we expect this flow to be triggered by the integrable deformation $\phi_{(1,3)}$ with no other operators contributing. It enters along the $\phi_{(1,3)}$ direction being the only irrelevant operator of $\cM(6,5)$ having the correct commutation relations with the lines $\left\{\one\comma\cL_{(1,2)}\comma\cL_{(1,3)}\comma\cL_{(1,4)}\comma\cL_{(4,1)} \right\}$, same RG invariants, and being the respective flow of the preserved lines in the UV.

\item {\large {$\boldsymbol{\cM(7,6) \xrightarrow{\phi_{(3,2)}} \cM(9,4)}$}.} This flow triggered by the deformation $\phi_{(3,2)}$ preserves the non anomalous $\bZ_2$ line $\cL_{(1,5)}$. The operator $\phi_{(3,3)}$ is generated along the flow and it enters along the $\phi_{(2,1)}$ direction. One also predicts flows to all the endpoints in the flows sitting on the series $\cM(9,4)\xrightarrow{\phi_{(1,5)}} \cM(7,4) \xrightarrow{\phi_{(1,4)}} \cM(8,3)$, and $\cM(8,5)\xrightarrow{\phi_{(1,2)}}\cM(8,3)$,  $\cM(9,4)\xrightarrow{\phi_{(1,5)}} \cM(7,4) \xrightarrow{\phi_{(1,4)}} \cM(5,4)$ {In this case the NLIE   only describe the flow to $\cM(5,4)$ }

\item {{\large $\boldsymbol{\cM(14,3) \xrightarrow{\phi_{(1,9)}} \cM(10,3)}$.}} This flow is triggered by $\phi_{(1,9)}$ and allows for generating along it all the $\phi_{(1,2k+1<9)}$. It preserves the not-anomalous $\bZ_2$ symmetry of the model. One also finds all the flows to the end point of the sequence $\cM(14,3)\to \cM(10,3)\to \cM(8,3)\to \cM(5,4) \to \cM(4,3)$, as well as the sequence to $\cM(14,3)\to \cM(10,3)\to\cM(7,4)$

\end{itemize}

\section{Appendix C: Recovering the known, integrable cases}\label{sec:reduction_to_known_cases}

The NLIEs {in eq. (7) of the main text} with kernels given {in eq. (9) there}, are conjectured to encode the spectrum of a very wide class of massless flows. Amongst these -- in particular, in the class of {Tanaka-}Nakayama's flows ({eq. (13) of the main text}) -- are the more familiar $\phi_{(1,3)}$, $\phi_{(1,5)}$, and $\phi_{(1,2)}$ flows, studied since the '90s in the literature \cite{Fendley:1993wq,Fendley:1993xa,Zamolodchikov:1994za,Fioravanti:1996rz,Feverati:1999sr,Dorey:2000zb}. It is straightforward to check that the choice $\kappa = 3$ yields the same kernels considered in \cite{Dorey:2000zb}. Indeed it is not difficult to see that the parameter $\mu$ in eq.\ (13) { of the main text} is related to $\kappa$ as $\mu = \kappa/z_{(1,2\mu + 1)} - 1$. Since $z_{(1,2\mu + 1)}$ can be either $1$ or $2$ for $\mathbb{Z}_2$ even or odd operators, respectively, we have the two options $\mu = 2$ or $\mu = 1/2$, corresponding, respectively, to $\phi_{(1,5)}$ and $\phi_{(1,2)}$ deformations.

Recovering $\phi_{(1,3)}$ is slightly more subtle\footnote{We thank Roberto Tateo for suggesting this calculation.}. In fact, one would need to set $\kappa = 2$ in the expressions of the kernels in {eq. (9) of the main text}, which is not immediately possible, since for $\kappa = 2$ the Fourier image of $\phi(\theta)$ is not integrable on $\mathbb{R}$. We need to proceed more carefully. We notice that
\begin{equation}
    \lim_{\kappa\to2}\hat{\phi}(\omega) = 2 \hat{\phi}_{\rm Z}(\omega) - 1\;,\qquad \lim_{\kappa \to2}\hat{\chi}(\omega) = 2 \hat{\chi}_{\rm Z}(\omega)\;,
\end{equation}
where $\hat{\phi}(\omega)$ and $\hat{\chi}(\omega)$ are the Fourier images of the kernels {in eq. (9) of the main text}. The Fourier images $\hat{\phi}_{\rm Z}(\omega)$ and $\hat{\chi}_{\rm Z}(\omega)$ are those used in \cite{Zamolodchikov:1994uw,Fioravanti:1996rz,Feverati:1999sr} to describe the $\phi_{(1,3)}$ massless flows. Then we find
\begin{equation}
    \lim_{\kappa \to 2} \phi(\theta) = 2\phi_{\rm Z}(\theta) - \delta(\theta) \;,\qquad \lim_{\kappa \to 2} \chi(\theta) = 2 \chi_{\rm Z}(\theta)\;.
\end{equation}
At the level of the NLIEs {in eq. (7) of the main text}, we take the limit $\kappa\to2$ and find
\begin{equation}
    \begin{split}
        f_{\rm R,Z}(\theta) &= \ri \alpha' - \ri \frac{r}{2} e^{\theta} - 2\sum_{\sigma=\pm} \sigma \intop_{\mathcal{C}_{\rm s}^{\sigma}} d\theta'\,\Big[\phi_{\rm Z}(\theta - \theta')L_{\rm R,Z}^{-\sigma}(\theta') + \chi_{\rm Z}(\theta - \theta')L_{\rm L,Z}^{\sigma}(\theta')\Big] + \\
        & + \sum_{\sigma=\pm} \sigma \intop_{\mathcal{C}_{\rm s}^{\sigma}} d\theta'\, \delta(\theta - \theta')L_{\rm R,Z}^{-\sigma}(\theta')\;, \\
        f_{\rm L,Z}(\theta) &= -\ri \alpha' - \ri \frac{r}{2} e^{-\theta} + 2\sum_{\sigma=\pm} \sigma \intop_{\mathcal{C}_{\rm s}^{\sigma}} d\theta'\, \Big[ \phi(\theta - \theta')L_{\rm L,Z}^{\sigma}(\theta') + \chi(\theta - \theta')L_{\rm R,Z}^{-\sigma}(\theta') \Big] + \\
        & - \sum_{\sigma=\pm} \sigma \intop_{\mathcal{C}_{\rm s}^{\sigma}} d\theta'\, \delta(\theta - \theta')L_{\rm L,Z}^{\sigma}(\theta')\;.
    \end{split}
    \label{eq:massless_NLIEs_towards_Z}
\end{equation}
Now, from the definition of the functions $L^\pm$ we have the identity
\begin{equation}
    \sum_{\sigma=\pm} \sigma \intop_{\mathcal{C}_{\rm s}^{\sigma}} d\theta'\, \delta(\theta - \theta')L_{\rm R,Z}^{-\sigma}(\theta') = - \intop_{-\infty}^{\infty} d\theta'\, \delta(\theta - \theta') f_{\rm R,Z}(\theta') = - f_{\rm R,Z}(\theta)\;,
\end{equation}
where we used the Cauchy theorem. A similar manipulation can be performed for the term with subscript L. Now, a slight shuffling of the furniture in \eqref{eq:massless_NLIEs_towards_Z}, yields
\begin{equation}
    \begin{split}
        f_{\rm R,Z}(\theta) &= \ri \frac{\alpha'}{2} - \ri \frac{r}{4} e^{\theta} - \sum_{\sigma=\pm} \sigma \intop_{\mathcal{C}_{\rm s}^{\sigma}} d\theta'\,\Big[\phi_{\rm Z}(\theta - \theta')L_{\rm R,Z}^{-\sigma}(\theta') + \chi_{\rm Z}(\theta - \theta')L_{\rm L,Z}^{\sigma}(\theta')\Big]\;, \\
        f_{\rm L,Z}(\theta) &= -\ri \frac{\alpha'}{2} - \ri \frac{r}{4} e^{-\theta} + \sum_{\sigma=\pm} \sigma \intop_{\mathcal{C}_{\rm s}^{\sigma}} d\theta'\, \Big[ \phi(\theta - \theta')L_{\rm L,Z}^{\sigma}(\theta') + \chi(\theta - \theta')L_{\rm R,Z}^{-\sigma}(\theta') \Big]\;,
    \end{split}
\end{equation}
which coincides precisely with the equations in \cite{Zamolodchikov:1994uw,Fioravanti:1996rz,Feverati:1999sr}, provided that we rescale $\alpha'$ and $r$ by a factor $2$.

\section{Appendix D: Chain of flows from $\boldsymbol{\cM(13,4)}$}
\begin{figure}[h]
    \centering
    \includegraphics[width=0.5\linewidth]{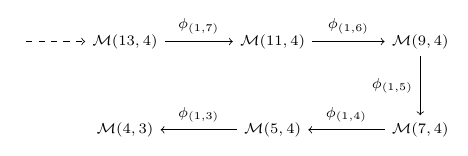}
    \caption{Chain of flows from $\cM(13,4)$ to Ising. We only report in the figure the most relevant deformation triggering the flow. }
    \label{fig:flowm413}
\end{figure}
We analyse here {the NLIE solutions} for the chain of flows\footnote{{This is a fragment of a semi-infinitely long sequence comprising  $\cM(4k+1,4)\xrightarrow{\phi_{1,2k+1}} \cM(4k-1,4)$ and $\cM(4k-1,4)\xrightarrow{\phi_{1,2k}} \cM(4k-3,4)$. Here, for concreteness, we chose to truncate it at $\cM(13,4)$.}} reported in  Figure \ref{fig:flowm413}, starting with $\cM(13,4)$ and landing ultimately on the Ising model $\cM(4,3)$. 
Using the procedure discussed in the main text we can split this flow in 
\begin{itemize}
    \item $\cM(13,4) \to \cM(11,4)$. The most relevant operator triggering the flow is $\phi_{(1,7)}$ preserving the TY$_2$ fusion category generated by the identity line, the invertible non-anomalous $\bZ_2$ line $\cL_{(3,1)}$ and   $\cL_{(2,1)}$ with $d_{(2,1)} = \sqrt{2}$. The spin content of the defect Hilbert space is:
    \begin{equation}
        s_{(2,1)} = \bZ \pm \left\{ \frac{1}{16}\comma \frac{7}{16} \right\}\comma \qquad s_{(3,1)} = \bZ \pm \left\{ 0\comma  \frac{1}{2} \right\}
    \end{equation}
    In the IR the flow enters in the $\phi_{(1,7)}$ direction in $\cM(11,4)$. 
    Furthermore the same category of lines is preserved by the entire tower $\{\phi_{(1,3)}, \phi_{(1,5)}, \phi_{(1,7)}\}$, that are therefore  dynamically generated along the  $\phi_{(1,7)}$ flow. 
The operators $\phi_{(1,2)},\phi_{(1,4)}, \phi_{(1,6)}$ commute only with the $\cL_{(1,3)}$, so they cannot be generated by $\phi_{(1,7)}$ flow. 
\item  $\cM(11,4) \to \cM(9,4)$. 
This flow is interesting because it explicitly break the Kramers-Wannier duality defect (this was also observed in \cite{Tanaka:2024igj}). The perturbations are $\phi_{(1,2)}, \phi_{(1,4)}, \phi_{(1,6)}$, in the IR $\phi_{(1,6)}$ arrives on $\phi_{(1,6)}$.
\item $\cM(9,4) \to \cM(7,4)$. Both  UV and IR have anomalous TY$_2$ category having  quantum dimensions $(1, -\sqrt{2}, 1)$. This subalgebra is preserved by $\phi_{(1,5)},\phi_{(1,3)}$ in the UV, with the former being the most relevant. They enter the IR in the $\phi_{(1,5)}, \phi_{(1,3)}$ directions. The flow is integrable, and single field with $\phi_{(1,5)}$.
\item  $\cM(7,4) \to \cM(5,4)$. Same feature as before: they break KW duality. As in $\cM(7,4)$ is anomalous, and in $\cM(5,4)$ is not. The most relevant operator in the UV is $\phi_{(1,4)}$, while $\phi_{(1,2)}$ also preserves $\bZ_2$. They enter the IR along the $\phi_{(1,2)}$ direction.
\item $\cM(5,4) \to \cM(4,3)$. This is the famous Zamolodchikov tri-critical Ising to Ising flow. The only allowed perturbing operator in the UV is $\phi_{(1,3)}$, and the flows enters the IR along the $T\bar{T}$ direction.

\end{itemize}

We numerically analysed the flows belonging to the chain in Figure \ref{fig:flowm413}. We fitted the numerical data against the expected UV and IR behaviours in eqs.\ (14) and (15) of the main text. We found excellent agreement with the predicted behaviour in all cases except the IR of $\cM(7,4) \to \cM(5,4)$, in which the numerical data was too unstable to be reliable. More quantitatively, the goodness-of-fit data is collected in Table \ref{tab:GoF_4_13_chain}. Note that all fits were performed with a CPT series truncated at 10 terms (bulk energy coefficient excluded).
\begin{table}[h!]
    \centering
    \begin{tabular}{| c |c|c|c|c|}
        \hline Flow & UV $\chi^2_{\rm red}$ & UV d.o.f. & IR $\chi^2_{\rm red}$ & IR d.o.f. \\ \hline
        $\mathcal{M}{(13,4)} \to \mathcal{M}{(11,4)}$ & 0.935364 & 39 & 1.69241 & 59 \\ \hline
        $\mathcal{M}{(11,4)} \to \mathcal{M}{(9,4)}$ & 1.47083 & 59 & 0.988184 & 58 \\\hline
        $\mathcal{M}{(9,4)} \to \mathcal{M}{(7,4)}$ & 1.02055 & 58 & 1.02304 & 58 \\\hline
        $\mathcal{M}{(7,4)} \to \mathcal{M}{(5,4)}$ & 5.21225 & 51 & 56439.4 & 38 \\\hline
        $\mathcal{M}{(5,4)} \to \mathcal{M}{(4,3)}$ & 1.16287 & 30 & 0.842416 & 48 \\ \hline
    \end{tabular}
    \caption{Quantitative, goodness-of-fit data of the behaviours ({eqs. (14) and (15) of the main text}) for the numerical data obtained by standard iteration of the NLIEs  for the chain of flows in Figure \ref{fig:flowm413}.}
    \label{tab:GoF_4_13_chain}
\end{table}

\begin{figure}
    \centering
    \includegraphics[width=1\linewidth]{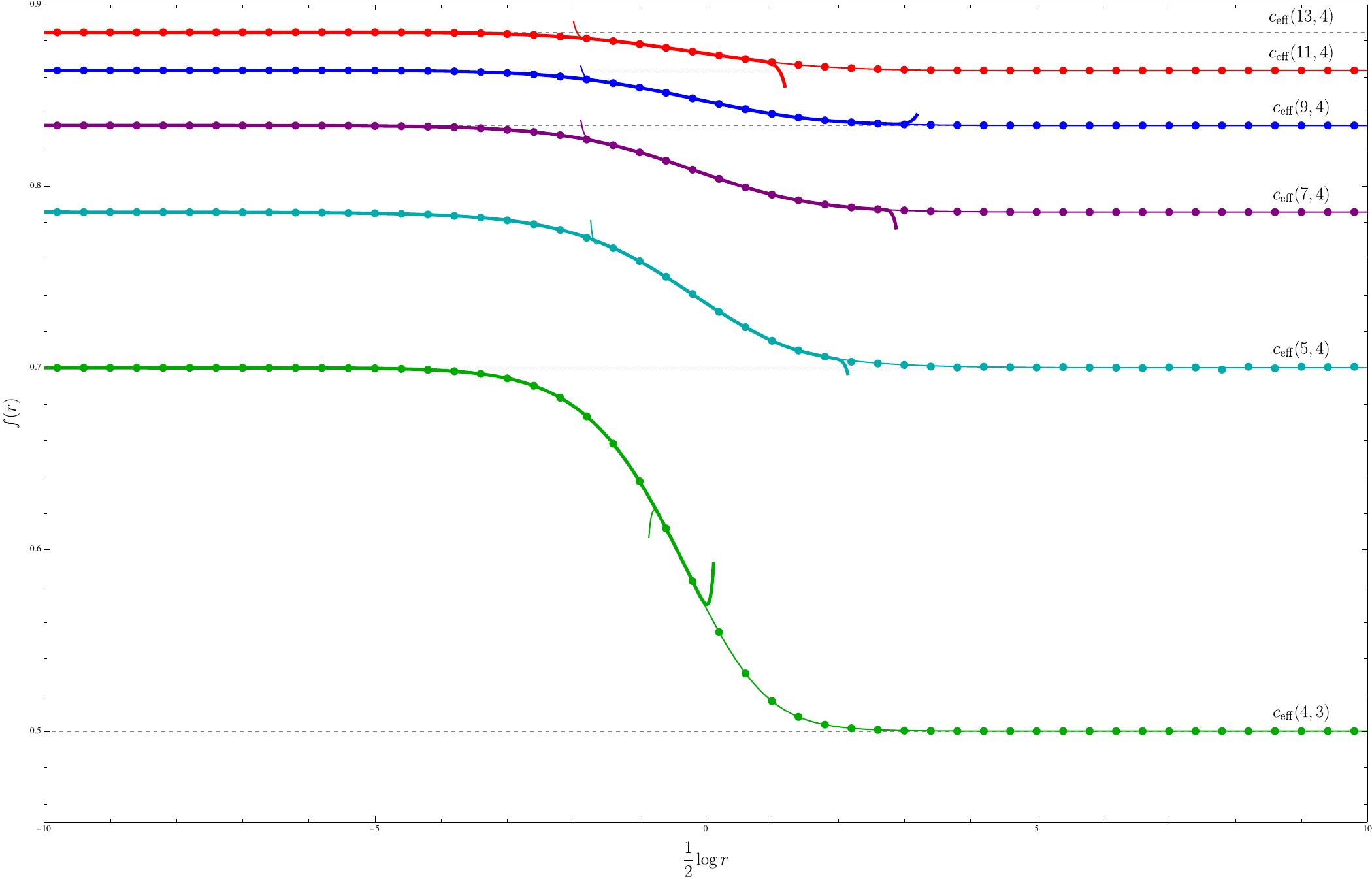}
    \caption{Plot of the scaling functions $f(r)$ for the chain of flows in Figure \ref{fig:flowm413}. The dots represent numerical data obtained by standard iteration of the NLIEs in {eq. (7) of the main text}. The thick and thin lines were obtained by fitting the numerical data against, respectively, the expected UV and IR behaviours ({eqs. (14) and (15) of the main text)}.}
    \label{fig:chain_plot}
\end{figure}

The results are represented graphically in Figure \ref{fig:chain_plot}.

